\documentclass[conference]{IEEEtran}
\IEEEoverridecommandlockouts
\usepackage{cite}
\usepackage{amsmath,amssymb,amsfonts}
\usepackage{algorithmic}
\usepackage{graphicx}
\usepackage{textcomp}
\usepackage{xcolor}
\usepackage{booktabs}
\usepackage{multirow}
\usepackage{multicol}
\usepackage{etoolbox}
\usepackage{hyperref}
\makeatletter
\patchcmd{\@makecaption}
  {\scshape}
  {}
  {}
  {}
\makeatletter
\patchcmd{\@makecaption}
  {\\}
  {.\ }
  {}
  {}
\makeatother
\def\tablename{Table}
\renewcommand{\figurename}{Figure}

\begin{document}

\title{Hard Exudate Segmentation Supplemented by Super-Resolution with Multi-scale Attention Fusion Module
}

\author{
\IEEEauthorblockN{Jiayi Zhang$^{1,\#}$, Xiaoshan Chen$^{1,\#}$, Zhongxi Qiu$^{2,\#}$, Mingming Yang$^{3}$,Yan Hu$^{1,*}$, Jiang Liu$^{1,2,*}$}
\IEEEauthorblockA{1 Department of Computer Science and Engineering, \\Southern University of Science and Technology, Shenzhen, 51805, Guangdong, China}
\IEEEauthorblockA{2 Research Institute of Trustworthy Autonomous Systems\\Southern University of Science and Technology, Shenzhen, 51805, Guangdong, China}
\IEEEauthorblockA{3 Department of Ophthalmology, Shenzhen People's Hospital, Guangdong, China}

\thanks{* Corresponding Author}
\thanks{\# These authors contributed equally to this work and should be considered co-first authors}
\thanks{Coreesponding To:\{huy3,liuj\}@sustech.edu.cn}
}

\maketitle

\begin{abstract}
Hard exudates (HE) is the most specific biomarker for retina edema. Precise HE segmentation is vital for disease diagnosis and treatment, but automatic segmentation is challenged by its large variation of characteristics including size, shape and position, which makes it difficult to detect tiny lesions and lesion boundaries. Considering the complementary features between segmentation and super-resolution tasks, this paper proposes a novel hard exudates segmentation method named SS-MAF with an auxiliary super-resolution task, which brings in helpful detailed features for tiny lesion and boundaries detection. Specifically, we propose a fusion module named Multi-scale Attention Fusion (MAF) module for our dual-stream framework to effectively integrate features of the two tasks. MAF first adopts split spatial convolutional (SSC) layer for multi-scale features extraction and then utilize attention mechanism for features fusion of the two tasks. Considering pixel dependency, we introduce region mutual information (RMI) loss to optimize MAF module for tiny lesions and boundary detection. We evaluate our method on two public lesion datasets, IDRiD and E-Ophtha. Our method shows competitive performance with low-resolution inputs, both quantitatively and qualitatively. On E-Ophtha dataset, the method can achieve $\geq3\%$ higher dice and recall compared with the state-of-the-art methods.

\end{abstract}


\section{Introduction}
Diabetic retinopathy (DR) is one of the vision defects diseases that leads to blindness. Hard exudate (HE) is one of the most prevalent clinical signs of retinopathy. It is also the most specific biomarker for retinal edema that is the primary cause of vision damage in non-proliferative DR \cite{Chew1996AssociationOE,doi:10.7326/0003-4819-116-8-660, SANCHEZ2009650}. Thus, segmenting hard exudate is significant for obtaining a timely diagnosis and treatment of DR, which can inhibit the progression of the disease and reduce the probability of blindness. Conventional segmentation methods require experienced ophthalmologists to discriminate the abnormal area using an ophthalmoscope, which is time-consuming and exhausting. Thus, the automatic hard exudates segmentation method plays a vital role in assisting ophthalmologists and improving the effectiveness of the diagnosis process\cite{Zhang2022}. However, several challenges exist in hard exudate segmentation, which hinders us from obtaining high-accuracy segmentation results.

\begin{figure}[htbp]
    \centering
    \includegraphics[width=0.34\textwidth]{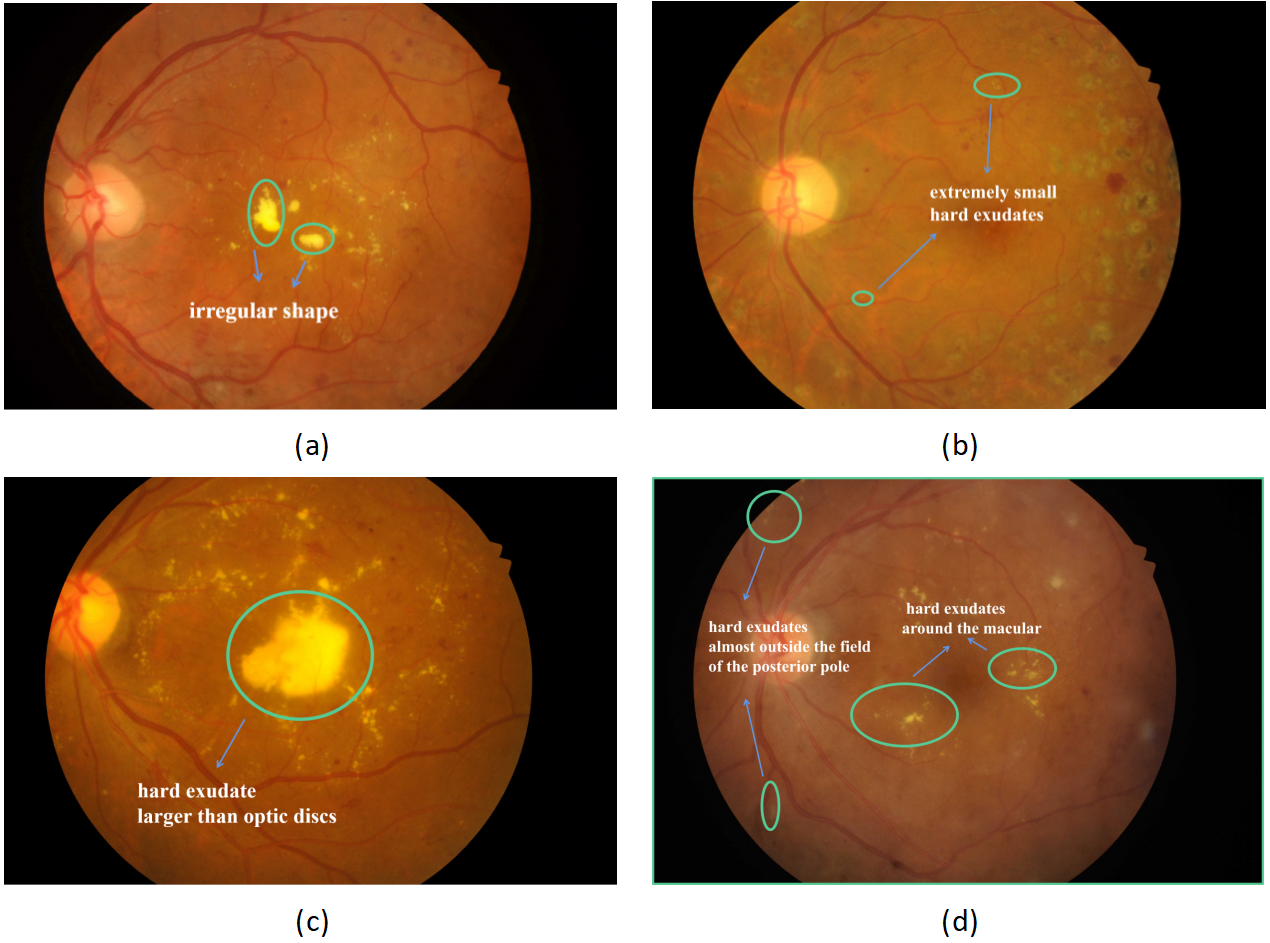}
    \caption{Examples of some difficulties in hard exudates segmentation, some hard exudates marked with green circles.}
    \label{fig:difficulties}
\end{figure}

There is no fixed character for hard exudates, compared with other retinal segmentation targets, such as optic discs and retinal vessels in color fundus images. As shown in Figure \ref{fig:difficulties}, the shapes of hard exudates are irregular, and the sizes can be larger than optic discs, or smaller than tiny vessels, as marked by green circles in Figure \ref{fig:difficulties}(a), (b) and (c). The positions of hard exudates are also uncertain, around the macular or almost outside the field of the posterior pole, as marked by green circles in Figure \ref{fig:difficulties} (d). The above characteristics result in some problems, like being unable to detect tiny lesions, class imbalance and difficulty in segmenting the lesion boundary precisely, which decrease the segmentation accuracy greatly.

To address the above challenges, researchers dedicated much effort. For example, LWENet\cite{guo2019lightweight} pretrains model with a DR classification branch to capture discriminative features for lesion segmentation. Fusion of high-level features and low-level features are employed in  FCRN \cite{mo2018exudate} and FFU-Net\cite{xu2021ffu} to improve the precision of boundary localization and detect small-size object. In addition, class-balanced cross entropy(CBCE) loss \cite{xie2015holistically} or variants of CBCE loss\cite{mo2018exudate}\cite{lin2017focal} are adopted in the above algorithms to solve the problem of class imbalance. However, most of the present algorithms cannot deal with some tiny hard exudates with fuzzy boundaries, as shown in Figure \ref{fig:difficulties}. The main reason is that the algorithm cannot provide high-resolution lesion features only based on low-resolution input, as limited by the computation abilities of some hospital machines.

In this paper, we try to improve the accuracy of hard exudate segmentation by precise boundary and tiny lesion localization. Super-resolution outputs high-resolution results only based on low-resolution input, which can solve the limitation of computation in the medical image segmentation application. Furthermore, it can provide detailed features to upsample images, and the features are supposed to be helpful for the segmentation task\cite{dai2016image,rad2019srobb,lei2019simultaneous,rad2020benefiting,DBLP:journals/corr/abs-2005-06382}. Thus, we propose to introduce super-resolution as an auxiliary task of hard exudate segmentation. Purely fusing features from super-resolution and segmentation cannot fully use the advantages of super-resolution, thus we propose Multi-scale Attention Fusion (MAF) module to effectively determine the boundary and tiny lesions. Moreover, considering the pixel dependency, we utilize a combination of CBCE loss and region mutual information(RMI)\cite{NEURIPS2019_a67c8c9a} loss to optimize the segmentation results to further improve the accuracy of lesion boundary. 

In summary, the main contributions of this paper are as follows:

$\bullet$We propose a novel segmentation method named SS-MAF combined with super-resolution features, which improves the hard exudates segmentation accuracy by detecting more tiny lesions and locating boundary precisely.

$\bullet$We propose a fusion module named Multi-scale Attention Fusion (MAF) module to improve the localization of tiny lesions and lesion boundaries, and the module integrates multiple receptive fields and attention mechanism to effectively utilize the super-resolution features.

$\bullet$We conduct experiments on two public datasets, IDRiD and E-Ophtha, and our method achieves competitive segmentation performance with low-resolution inputs on hard exudate segmentation. On the E-Ophtha dataset, our algorithm can achieve $\geq3\%$ imrpovement in dice and recall compared with the state-of-the-art algorithms.

\section{Related Works}
In this section, we introduce related works from three aspects, including hard exudate segmentation, single image super-resolution and dual-stream super-resolution semantic segmentation. 

\subsection{Hard exudates segmentation}
Researchers dedicated much effort to improving the accuracy of hard exudate segmentation, as it is fundamental to DR diagnosis. They proposed many strategies to obtain accurate segmentation results, such as pretrained encoder, multi-scale feature fusion, and so on. For example, FCRN \cite{mo2018exudate} and HED \cite{xie2015holistically} adopt pretrained ResNet-50 \cite{He2016DeepRL} and VGG16\cite{Simonyan2015VeryDC}, respectively, to capture rich semantic features. Different multi-scale feature fusion modules are proposed in LWENet\cite{guo2019lightweight}, PMC-Net\cite{He2022ProgressiveMC} , L-Seg\cite{guo2019seg} and FFU-Net\cite{xu2021ffu} to precisely localize boundaries and tiny lesions. Moreover, researchers also proposed some losses to optimize networks, such as CBCE loss\cite{xie2015holistically}, variants of CBCE\cite{mo2018exudate, lin2017focal}, and so on. However, present algorithms cannot segment tiny lesion or precise lesion boundary.

\subsection{Single Image Super-Resolution}
In the medical fields, low spatial resolution usually leads to errors in post-processing. Recently, researchers propose some deep-learning based super-resolution methods\cite{zhao2019channel, shi2018super} to achieve promising results. Park \textit{et al.}\cite{Park2018ComputedTS} propose a modified U-Net for 2D brain Computed Tomography(CT) images while Liu \textit{et al.}\cite{liu2018fusing} design a multi-scale features fusion module. Moreover, Zhang \textit{et al.}\cite{zhang2018fast} propose a fast medical image super-resolution method for retinal images based on a sub-pixel convolution layer. 

\subsection{Dual-Stream Super-Resolution Semantic Segmentation}
Reconstructed high-resolution images can be used as a source of information supplementation for semantic segmentation. Comparable results can be obtained from low-resolution image by introducing super-resolution task, which greatly alleviates the dilemma of limited computational resources. In the field of natural scene, Wang et.al\cite{2020Dual} propose a dual-stream model with a feature affinity module for computing the similarity of the segmentation and the super-resolution features. In the field of medical images, researchers focus on fusing features of the two streams from different aspects. SuperVessel\cite{hu2022supervessel} employs two specific modules to enhance the features of the interested segmentation area for vessel segmentation. Sang et.al\cite{2021Super} propose an infection edge detection guided region mutual information loss for COVID-19 CT segmentation. A patch-free 3D segmentation method is \cite{2021Patch} proposed with a Self-Supervised Guidance Module and a Task-Fusion Module to explore the connection between those two streams for 3D image segmentation. 

\section{Method}
The overall structure for our method is shown in Figure \ref{fig:whole}. Similar to other dual-stream methods, our SS-MAF model consists of one semantic segmentation stream, one super-resolution stream, and one fusion module named Multi-scale Attention Fusion(MAF) module. In the following, we first introduce the dual-stream learning, then describe the MAF module in detail. Finally, we discuss the design of the objective function.

\subsection{Dual-stream learning}
\begin{figure*}[!htbp]
    \centering
    \includegraphics[width=0.75\textwidth]{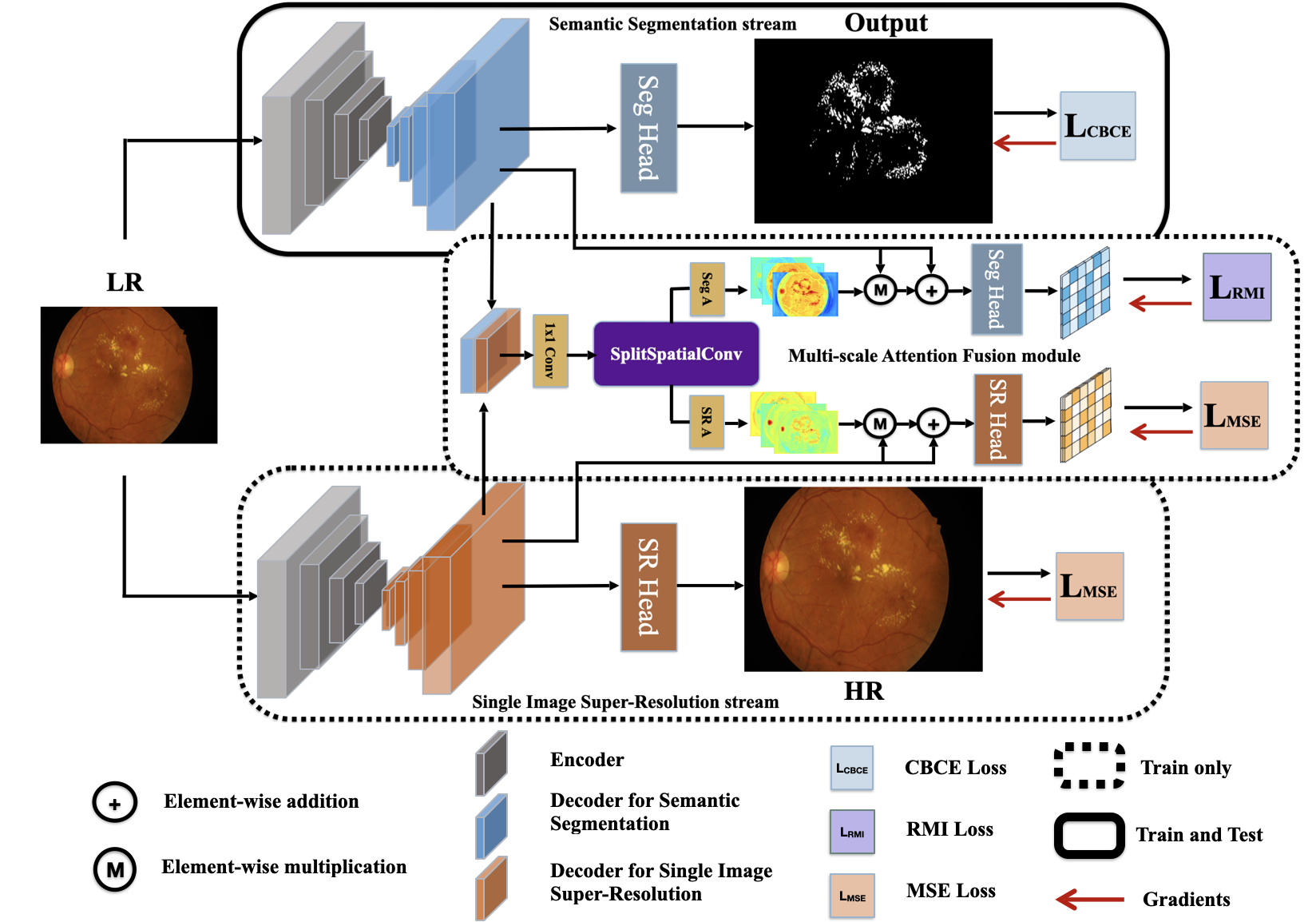}
    \caption{The figure shows the overall structure of our proposed method. The above part represents semantic segmentation stream while the part below represents single image super-resolution stream. Between the two streams is the fusion module that efficiently fuses features of the two streams and extracts their mutual information to obtain more precise segmentation results. We mention that only the segmentation stream is used for testing, which significantly reduces the time and computational resources needed for testing.}
    \label{fig:whole}
\end{figure*}

Semantic segmentation and super-resolution are two independent tasks, however, they can improve their performance by learning from each other\cite{dai2016image,rad2019srobb,lei2019simultaneous,rad2020benefiting,DBLP:journals/corr/abs-2005-06382}. One simple but effective way to combine super-resolution and semantic segmentation is dual-stream learning. For our dual-stream network as Figure \ref{fig:whole} shows, one shared encoder encodes the input image to encoded features, then two task-specific decoders are adopted to decode the encoded features to task-dependent features based on different aims of the tasks. Finally, the task-specific head can generate the output of the task according to the task-dependent features.

Specifically, given an image $x$ of size $H \times W$, and an upsampling factor $N$, the image is first fed into the shared encoder to obtain the encoded features $F_{en}$. Then $F_{en}$ forward to the decoder of super-resolution and semantic segmentation respectively, and the super-resolution dependent features $F_{SR}$ and semantic segmentation dependent features $F_{Seg}$ are obtained separately. According to the $F_{Seg}$, the segmentation head(Seg Head), implemented by one 1$\times$1 convolution and one interpolation operation, generates the segmentation probability map of size $(H \times N) \times (W \times N)$.  The super-resolution head(SR Head) reconstructs the high-resolution image of size $(H \times N) \times (W \times N)$ based on $F_{sr}$ at the same time, and we use the ESPCN\cite{Shi_2016_CVPR} as the SR head. The whole process can be formulated as follows:
\begin{gather}
        F_{en} = \mathrm{E}(x) \\
        F_{Seg} = D_{Seg}(F_{en}) \\
        F_{SR} = D_{SR}(F_{en}) \\
        O_{Seg} = H_{Seg}(F_{Seg}) \\
        O_{SR} = H_{SR}(F_{SR})
\end{gather}
where E is the shared encoder, $D_{Seg}$ is the decoder of segmentation stream, $D_{SR}$ is the decoder of super-resolution stream, $H_{Seg}$ and $H_{SR}$ are the task head of segmentation and super-resolution respectively. In this paper, we build the dual-stream model based on the U-Net, and in section \ref{ablation:part} we conduct the ablation study to verify the effectiveness of each part.

\subsection{Multi-scale Attention Fusion Module}
\begin{figure}[htbp]
    \centering
    \includegraphics[width=0.4\textwidth]{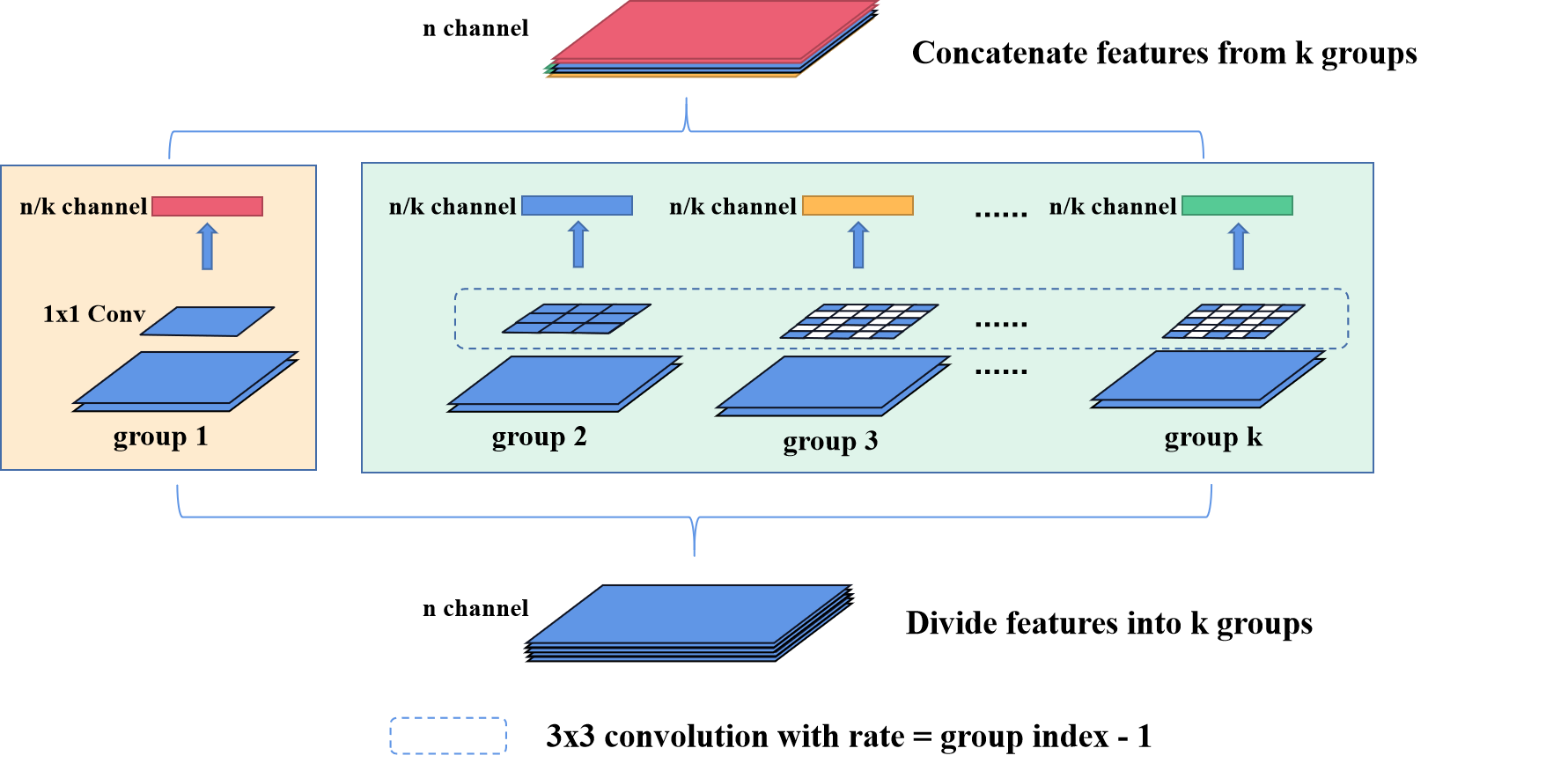}
    \caption{Detailed structure of split spatial convolutional (SSC) layer \textbf{(Please zoom in for a better view.)}}
    \label{fig:fusion}
\end{figure}

The super-resolution can bring information that helps locate the small lesions and boundaries, but some noise is introduced simultaneously due to its property. To alleviate this problem, we propose a module named Multi-scale Attention Fusion(MAF) that can make full use of the fusion features of two tasks by integrating different scale information of fused features.
As shown in Figure \ref{fig:whole}, the features of two tasks are firstly concatenated, then one $1\times1$ convolution is utilized to align the features of two tasks and map the fusion features to the hidden feature space of $d$ dimension, in this paper we set $d=32$. Then the fused features are fed into the split spatial convolutional(SSC) layer, and the detailed structure of SSC layer is shown in Figure \ref{fig:fusion}. The fused features are divided into $k$ groups on the basis of dimension, and for the first group, one $1\times1$ convolution is used to keep the information of the current scale. For the other k-1 groups, $3\times3$ convolutions with different dilation rates are employed to capture information on different scales, and the dilation rate is set to index number of the group minus 1. In this way, the layer can gain multi-scale information for the task fusion features, and the different groups can learn different ranges of information around the pixel that can help recognize the boundary of objects through the backpropagation gradient. Features of each convolution group are concatenated together, and batch normalization is added to integrate the information of each group. The formulation of this process is as follows:
\begin{equation}
    F_{fusion} = SSC(\theta(C(F_{Seg}, F_{SR})))
\end{equation}
where $\theta$ is the convolution operation and $C$ represents the concatenate operation.
Finally, two separate attention modules for segmentation (Seg A) and super-resolution (SR A) are used to generate the attention weights respectively. The attention weights are utilized to re-weight the features of the corresponding task, and the corresponding task head processes the features after re-weight operation to get the fusion task output that is independent of the dual-stream output. The whole process can be described as:
\begin{gather}
    W_{Seg} = \delta(\theta_{Seg}(F_{fusion})) \\
    W_{SR} = \delta(\theta_{SR}(F_{fusion})) \\
    O_{FuSeg} = H_{Seg}(W_{Seg}F_{Seg} + F_{Seg}) \\
    O_{FuSR} = H_{SR}(W_{SR}F_{SR} + F_{SR})
\end{gather}
where $\delta$ is sigmoid function, $\theta_{Seg}$ and $\theta_{SR}$ are the $1\times1$ convolution used in segmentation attention module and super-resolution attention module respectively. $H_{Seg}$ and $H_{SR}$ share weights with the task head in the corresponding task stream.
\subsection{Objective function}
The whole objective function consists of the loss of the three components in our model, the semantic segmentation stream, the super-resolution stream, and the MAF module. 

For the semantic segmentation stream, since the distribution of background pixels and hard exudates pixels is highly imbalanced, making the model easily ignore small hard exudates, we train the stream with the modified CBCE loss named Class-Balanced Softmax Cross-Entropy Loss\cite{8953804}. The formulation is as follows:
\begin{equation}
    L_{CBCE} = \frac{1}{C}\sum_{i=0}^{C}-\frac{1-\beta}{1-\beta^{n_i}}y_{i}\log(z_{i})
\end{equation}
where $C$ represents the number of classes, $\beta$ is a super parameter that is set to 0.9999 in our experiment, $n_i$ represents the number of pixels of class $i$ in the image, $y_i$ is the ground truth of class $i$, and $z$ is the softmax result of the output $O_{Seg}$.

For the super-resolution stream, we utilize the mean square error(MSE) as the objective function. The formulation is as follows:
\begin{equation}
    {L}_{MSE} =  \frac{1}{N}\sum_{i=0}^N(O_{SR_i} - HR_i)^2
\end{equation}
where $N$ is the number of pixels of the images, $O_{SR}$ is the output of  the super-resolution stream, $HR$ is the high-resolution target image, and $i$ represents the index of the pixel.

\label{section:loss}
For the MAF module, the aim is to make the interaction between two streams remit the noise brought by super-resolution stream and make the two streams learn from each other. The intuitive idea is that the module can generate a better result for each sub-stream, thus losses for both subtasks can be utilized here to constrain the module. However, the MAF module is desired to get information on the region for more accurate locating of the boundary. We then introduce one region-based loss named region mutual information(RMI) loss that takes the correlation among pixels in a nearby region into account\cite{NEURIPS2019_a67c8c9a}, and use this loss as the objective function for the fusion segmentation part. For super-resolution part in MAF, we follow the strategy of using the same loss of the subtask as the objective function, then the final loss for the MAF can be formulated as:
\begin{equation}
      {L}_{MAF} = {L}_{RMI}(O_{FuSeg},Y) + {L}_{MSE}(O_{FuSR},HR)
\end{equation}
where $O_{FuSeg}$ and $O_{FuSR}$ are the segmentation output and the super-resolution output from the MAF module, $Y$ is the ground truth of the segmentation task, and $HR$ is the targeted high-resolution image. 

Thus, our whole objective function can be described as follows:
\begin{equation}
    {L}_{total}={L}_{CBCE} + {L}_{MSE} + {L}_{MAF}
\end{equation}

\section{Experiments}
\subsection{Datasets}
We conduct our experiments on two public available retinal datasets with hard exudates annotation: IDRiD\cite{Porwal2018IndianDR} and E-Ophtha-EX\cite{Zhang2014ExudateDI}. IDRiD dataset consists of 81 color fundus images and pixel level annotations of retinal lesions. According to the official separation, 54 images are for training, and the rest 27 images are for testing. E-Ophtha dataset provides 82 retinal images, of which 47 images with hard exudates. In our experiments, for both datasets, we resize images to $720\times480$ as input and the output image size is set to $1440\times960$. 

\subsection{Evaluation Metrics}
The experimental results are evaluated on pixel-level metrics of the exudates class, including dice, intersection over union(IoU), recall, and area under the curve(AUC). AUC is the area under the curve of precision and recall obtained by varying the threshold from 0 to 1.

\subsection{Implementation Details}
All the experiments are run based on Pytorch framework and an NVIDIA RTX 2080TI graphics card. Models are trained with initial learning rate $init\_lr=0.01$ and poly learning rate strategy where the learning rate is calculated by $init\_lr\times(1-\frac{iter}{max\_iter})^{power}$ with $power=0.9$. Moreover, the mini-batch stochastic gradient (SGD) with the momentum 0.9 and the weight decay 0.0001 are applied. The training epoch is set to 300 for all experiments and batch size is set to 2 on IDRiD dataset while 1 on E-Ophtha dataset. 


\subsection{Ablation Study}

\noindent\textbf{Ablation study of different counterparts}
\label{ablation:part}
The effectiveness of our SS-MAF model is verified by performing an ablation study on different fractions. Experimental results and corresponding fractions of both datasets are demonstrated in Table \ref{tab:ablation}. Interp represents a simple upsampling operation to achieve the specified output size while SR and MAF constitute the super-resolution stream and the MAF module respectively. The results of adding a simple upsampling operation are barely the same as the baseline or even lower, while adding the super-resolution stream outperforms the baseline on both datasets. This demonstrates that the way of utilizing super-resolution stream as information supplement is valid. In addition, the results of simply conducting the segmentation task with super-resolution are not as satisfactory as the ones obtained by applying the MAF module, which significantly verifies the effectiveness of MAF Module.
\begin{table}[htbp]
    \centering
     \caption{Results for ablation study on IDRiD and E-Ophtha datasets($\mathrm{mean}$ of five times).}
    \resizebox{0.5\textwidth}{!}{
    \begin{tabular}{lcccccc}
    \hline
        \toprule
        Dataset&Model & Dice(\%) & IoU(\%) & Recall(\%) & AUC(\%) \\
        \midrule
        \multirow{4}{*}{IDRiD}&U-Net & $56.22$ & $39.11$ & $47.02$ & $84.79$ \\ 
        ~&U-Net+Interp & $56.62$  & $39.56$ & $45.4$ & $82.56$ \\ 
        ~&U-Net+Interp+SR & $57.44$& $40.36$ & $46.42$ & $82.49$ \\ 
        ~&U-Net+Interp+SR+MAF & $\mathbf{72.55}$ & $\mathbf{56.93}$ & $\mathbf{69.39}$ & $\mathbf{85.96}$ \\ 
        \midrule
         \multirow{4}{*}{E-Ophtha}&U-Net& $46.59$ & $30.39$ & $55.7$ & $79.69$ \\ 
        ~&U-Net+Interp & $44.99$ &  $29.15$ & $40.35$ & $73.62$ \\ 
        ~&U-Net+Interp+SR & $51.12$ & $34.36$ & $\mathbf{60.66}$ & $\mathbf{85.07}$ \\ 
        ~&U-Net+Interp+SR+MAF & $\mathbf{52.96}$ & $\mathbf{36.03}$ & $57.8$ & $84.42$ \\  
        \bottomrule
    \end{tabular}
    }
    \label{tab:ablation}
\end{table}

\begin{figure}[htbp]
    \centering
    \includegraphics[width=0.45\textwidth]{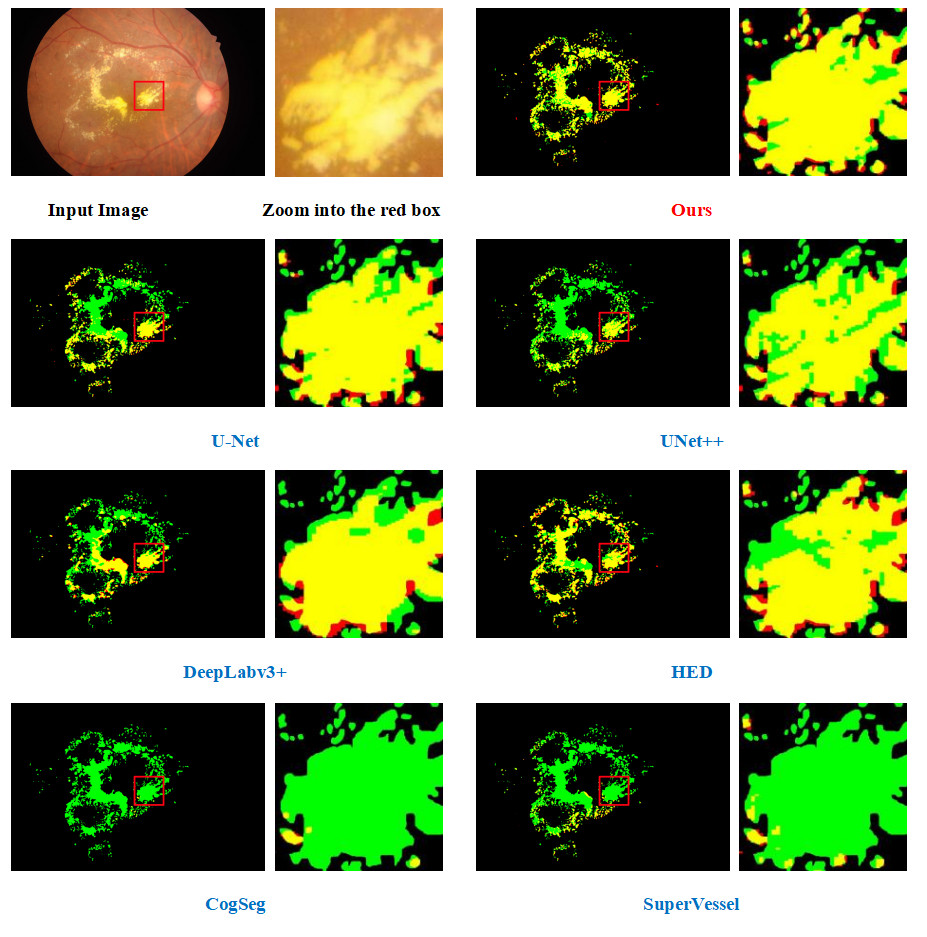}
    \caption{Segmentation results of different methods on IDRiD. Green area represents ground truth lesion parts while Red area represents output of specific method and yellow area is the area that the model predicts correctly.\textbf{(Please zoom in for a better view.)}}
    \label{fig:idrid_com}
\end{figure}

\subsection{Comparison with State-of-the-art methods}
On IDRiD dataset, we compare our SS-MAF model with U-Net\cite{ronneberger2015u}, DeepLabv3+\cite{chen2018encoder}, HED\cite{xie2015holistically}, UNet++\cite{Zhou2018UNetAN}, CogSeg\cite{2021Super}, SuperVessel\cite{hu2022supervessel}, LWENet\cite{guo2019lightweight}, FCRN\cite{mo2018exudate}.
Among them, the results of LWENet\cite{guo2019lightweight} and FCRN\cite{mo2018exudate} are cited from the original papers, which are obtained by having input images of size $1440\times960$, which is two times of ours, and batch size 2. All results are illustrated in Table \ref{tab:idrid_SOTA} and it shows that our method outperforms all methods with the same settings on IoU, dice, recall. Noticing that results of  LWENet\cite{guo2019lightweight} is obtained with a pretraining process, we further compare our result with their result  without pretraining, as shown in Table \ref{tab:idrid_SOTA}, SS-MAF can achieve competitive performance with a higher dice. Besides, as shown in Figure \ref{fig:idrid_com}, the visual results of our method have a better capability of localizing and distinguishing edges of exudate lesions than other mentioned methods. 
\begin{table}[htbp]
    \centering
     \caption{Results for comparison with state-of-the-art methods on IDRiD dataset($\mathrm{mean}$ of five times). LWENet$^\dagger$ and LWENet represent models with and without pretraining respectively.}
    \resizebox{0.5\textwidth}{!}{
    \begin{tabular}{lcccccc}
        \toprule
        Model & Dice(\%) & IoU(\%) & Recall(\%) & AUC(\%) \\
        \midrule
        U-Net & $62.28$ & $45.26$ & $52.67$ & $86.54$ \\ 
        DeepLabv3+& $58.21$ & $41.10$ & $47.81$ & $84.07$ \\ 
        HED & $68.96$ & $52.68$ & $65.5$ & $38.98$ \\ 
        UNet++ & $64.3$ &  $47.48$ & $56.28$ & $\mathbf{87.73}$ \\ 
        \midrule
        LWENet& ${69.68}$ & $-$ & ${69.84}$ & $-$ \\ 
        LWENet$^\dagger$&$\mathbf{78.15}$ & $-$ & $\mathbf{78.03}$ & $-$  \\ 
        FCRN & $64.12$ & $-$ & $68.62$ & $-$ \\ 
        \midrule
        CogSeg & $63.44$ &  $46.63$ & $55.56$ & $63.36$ \\ 
        SuperVessel & $51.23$ & $35.25$ & $38.4$ & $69.92$ \\ 
        SS-MAF(Ours) & $72.55$ & $\mathbf{56.93}$ & $69.39$ & $85.96$ \\ 
        \bottomrule
    \end{tabular}
    }
    \label{tab:idrid_SOTA}
\end{table}

\section{Conclusion}
In this paper, we propose a segmentation method named SS-MAF that effectively fuses the high-resolution features from a super-resolution stream to obtain exudate segmentation with precise boundaries and tiny lesions. Our work utilizes a fusion module with multi-scale receptive fields and the attention mechanism to accurately extract information about exudate lesions from the segmentation stream and the super-resolution stream. The module is trained by RMI loss to better localize the boundary lines of lesions considering region pixel dependency. The results of experiments on two public datasets illustrate the effectiveness and competitiveness of our work.

\section{Acknowledgement}
This work was supported in part by The National Natural Science Foundation of China(8210072776), National College Students' Innovation and Entrepreneurship Training Program(202214325007), Guangdong Provincial Department of Education (2020ZDZX3043), Guangdong Basic and Applied Basic Research Foundation(2021A1515012195).

\bibliographystyle{IEEEtran}
\bibliography{ref}

\end{document}